\documentclass[preprint]{aastex}

\newcommand{\lum}{\,{\rm erg\,s^{-1}}}

\begin{document}

\title{An X-ray Selected Galaxy Cluster at $z=1.11$ in the Rosat Deep
Cluster Survey\altaffilmark{1,}\altaffilmark{2,}\altaffilmark{3}}

\author{S.A.\ Stanford\altaffilmark{4}, Brad Holden\altaffilmark{4}} 
\affil{Physics Department, University of California Davis, Davis, CA 95616}
\email{adam@igpp.ucllnl.org, bholden@igpp.ucllnl.org}
 
\author{Piero\ Rosati} 
\affil{ESO -- European Southern Observatory, \\D-85748 Garching bei
M\"unchen, Germany} 
\email{prosati@eso.org}

\author{Peter R.\ Eisenhardt, Daniel Stern} 
\affil{Jet Propulsion Laboratory, California Institute of Technology, Pasadena, CA 91109}
\email{prme@kromos.jpl.nasa.gov, stern@zwolfkinder.jpl.nasa.gov}

\author{Gordon Squires}
\affil{California Institute of Technology, Pasadena, CA 91109}
\email{gks@astro.caltech.edu}

\and
\author{Hyron Spinrad}
\affil{Astronomy Department, University of California, 
Berkeley, CA  94720}
\email{spinrad@bigz.berkeley.edu}

\altaffiltext{1}{Based in part on observations obtained at the W.M.\ Keck
Observatory}
\altaffiltext{2}{Based in part on observations obtained at Palomar
Observatory}
\altaffiltext{3}{Based in part on observations obtained with the Chandra Observatory}
\altaffiltext{4}{Institute of
Geophysics and Planetary Physics, Lawrence Livermore National
Laboratory, Livermore, CA 94550}

\begin{abstract}
We report the discovery of an X-ray luminous galaxy cluster at $z =
1.11$.  RDCS J0910+5422 was selected as an X-ray cluster candidate in
the ROSAT Deep Cluster Survey on the basis of its spatial extent in a
Rosat PSPC image.  Deep optical and near-IR imaging reveal a red
galaxy overdensity around the peak of the X-ray emission, with a
significant excess of objects with $J-K$ and $I-K$ colors typical of
elliptical galaxies at $z \sim 1$.  Spectroscopic observations at the
Keck II telescope secured 9 galaxy redshifts in the range
$1.095<z<1.120$ yielding a mean cluster redshift of $<z>=1.106$.  Eight of
these galaxies lie within a 30 arcsec radius around the peak X--ray
emission.  A deep Chandra ACIS exposure on this field shows extended
X-ray morphology and allows the X-ray spectrum of the intracluster
medium to be measured.  The cluster has a bolometric luminosity $L_x =
2.48^{+0.33}_{-0.26}\times 10^{44}$ ergs s$^{-1}$, a temperature of
$kT = 7.2^{+2.2}_{-1.4}$ keV, and a mass within $r = 1$ Mpc of $7.0
\times 10^{14} M_{\sun}$ ($H_0=65$ km s$^{-1}$ Mpc$^{-1}$,
$\Omega_m = 0.3$, and $\Omega_\Lambda = 0.7$).  The spatial distribution of the
cluster members is elongated, which is not due to an observational
selection effect, and followed by the X-ray
morphology.  The X-ray surface brightness profile and the
spectrophotometric properties of the cluster members suggest that this
is an example of a massive cluster in an advanced stage of formation
with a hot ICM and an old galaxy population already in place at $z >
1$.

\end{abstract}


\section{Introduction}

The identification and study of distant galaxy clusters is of great
interest in current astronomical research.  As the largest
gravitationally bound structures in the universe, the properties and
histories of galaxy clusters are highly sensitive to the physics of
cosmic structure formation and to the values of the fundamental
cosmological parameters \citep{eke96,bahcall97}.  While clusters in
such well-defined samples as the Rosat Deep Cluster Survey
\citep[RDCS]{rosati98} have been used to constrain $\Omega_m$ and $\sigma_8$
\citep{borg01a}, the uncertainty in the relation between cluster mass
and measurables such as $L_x$ or $T_x$ limits the precision obtainable
in calculations of cosmological parameters to $\sim$50\%.  To improve
on this, a better understanding of the mass--$T_x$ relation and its
evolution is necessary.  To that end, we need to study in detail a
well-defined sample of clusters at high redshift with independent
measures of the cluster mass based on $T_x$ \citep{evrard96} and weak
lensing \citep{tyson90}.

Another equally important use of galaxy clusters lies in studying the
formation and evolution of galaxy populations.  Progress in
understanding early-type galaxy evolution in clusters is being driven
by the need to reproduce recent observational results indicating both
lower number fractions \citep{dressler97,couch98,dokkum00} {\it and}
strong homogeneity and slow evolution in the stellar populations of
ellipticals and S0s in moderate redshift clusters
\citep{aragon93,ellis97,sed98,pahre98,bender98,dokkum98a,rdp99,kelson00}.
A self-consistent explanation of these results can be achieved by
invoking an observational bias: the progenitors of the youngest,
low-redshift early-types drop out of samples constructed in high
redshift clusters.  \citet{vf01a} have suggested how 
morphological evolution at $z \lesssim 2$ coupled with star
formation at $z \sim 2-3$ ($\Omega_m = 0.3$, $\Lambda = 0.7$) can
explain the results cited above.  More complex semi-analytic models of
galaxy formation and evolution set in a CDM universe naturally predict
the morphological evolution that is a central tenet in the currently
fashionable paradigm of cluster galaxy evolution: ellipticals are
formed by mergers at $0.5 < z < 1.5$ of sub $L^\ast$ galaxies formed
at $2 < z < 3$ \citep{dokkum99}, and the Butcher-Oemler
effect is the result of spirals being converted into S0s at moderate
redshifts \citep{dressler97}.  For example, the models of
\citet{kf98} are able to match the small scatter in the
color-mag relation of the early-types even at $z \sim 1$, and the
$M/L$ evolution found by e.g.\ \citet{dokkum98b} can be
predicted \citep{dia01}.  However the accuracy as well as
the details of this scenario for galaxy evolution in clusters is yet
open to debate.  In the case of the early-types, the arguments are
based heavily on observations of only a few high-z clusters e.g.,
MS1054$-$03 at $z=0.82$ \citep{dokkum98b}.

Finding clusters at moderate redshifts has become almost routine using
serendipitous X-ray searches.  The advent of the ROSAT-PSPC, with its
unprecedented sensitivity and spatial resolution, enabled archival
searches for extended X-ray sources to become a very efficient method
to construct large, homogeneous samples of galaxy clusters out to 
$z\simeq 0.9$ \citep{rosati98,scharf97,collins97,vihk98}.  The
RDCS has shown no evidence of a decline in the space density of galaxy
clusters of X-ray luminosity $L_X\lesssim L_X^\ast$ over a wide
redshift range, $0.2<z<0.8$ (see Rosati 2000 for a recent
review), though the evolution of the bright end remains controversial.
The fact that the bulk of the X-ray cluster population is not evolving
significantly out to this large redshift increases the chances of
finding clusters at even higher redshifts, since $L_{X,z=0}^\ast$
clusters at $z>1$ ($\approx 4\times 10^{44}\lum$ in the [0.5-2.0] keV
band, roughly the Coma cluster) can be detected as extended X-ray
sources in deep ROSAT pointed observations, provided that the X-ray
surface brightness profile does not evolve significantly.

At fainter fluxes ($F_x < 3 \times 10^{-14}\ {\rm ergs\ cm^{-2}\
s^{-1}}$ in the RDCS) the identification of real clusters becomes more
difficult due to the increasing confusion and low signal-to-noise
ratio of the X-ray sources which makes it more difficult to
discriminate between point-like and extended sources.  Below $F_x < 2
\times 10^{-14}\ {\rm ergs\ cm^{-2}\ s^{-1}}$ the X-ray completeness
level can be as low as 50\%, and the spurious rate as high as 50\%.
But it is at these flux levels that the most distant clusters in the
survey are expected.  To improve the success rate of identifying very
high redshift clusters in the RDCS, we have been carrying out a
program of near-infrared imaging of faint unidentified X-ray
candidates.  Relative to the optical, near-IR imaging is advantageous
at high redshifts because the expected k-correction significantly dims
the dominant population of early-type cluster galaxies in even the
observed $I$-band for $z \gtrsim 1$.  \citet{s97} have shown that
optical-infrared colors can be used to considerably enhance the
contrast of high redshift cluster galaxies against the field galaxy population
and, at the same time, to obtain a useful estimate of the cluster
redshift.  The cluster found by \citet{s97}, RDCS J0848+4453 at
$z=1.27$, and its neighbor, RDCS J0849+4452 at $z=1.26$
\citep{rosati99}, have been used to push the study of evolution in the
colors and morphology of early-type galaxies beyond $z > 1$.  But such
studies continue to be severely limited by the dearth of such high
redshift clusters \citep{dick97,fabian01}.

In this paper, we describe the imaging and spectroscopic follow-up
observations of the extended X-ray source RDCS J0910+5422, which has led
to the discovery of a galaxy cluster at $z=1.106$.  Unless otherwise
stated, we adopt the parameters $H_0=65$ km s$^{-1}$ Mpc$^{-1}$,
$\Omega_m = 0.3$, and $\Omega_\Lambda = 0.7$.

\section{Observations}

\subsection{Optical and IR Imaging}

The extended X-ray source RDCS J0910+5422 was selected from a deep
ROSAT-PSPC observation; details of the selection procedure are given
in \citet{rosati99}.  The candidate field was observed in the
optical at the Palomar 5~m telescope with COSMIC, which contains a
TEK3 CCD that provides $0\farcs28$ pixels over a 9.5 arcmin field of
view.  An exposure of 2820~s was obtained in the gunn $i$-band on 18
February 1999 UT in non-photometric conditions with $1\farcs4$ seeing.
The field was observed again with COSMIC in the gunn $i$-band in
photometric conditions for 300~s on 2000 April 30.  5 standard stars
were observed on the latter night and used to calibrate the photometry
onto the Vega system.  The data were reduced using standard methods.
The second observation was used to calibrate the deeper,
nonphotometric image before the two reduced images were summed.

We obtained $J$ and $K_s$ imaging at the Palomar 5~m telescope
with the Prime-Focus Infrared Camera \citep{jarrett94}.  This
camera provides a $2\farcm1$ field of view with $0\farcs494$ pixels.
RDCS J0910+5422 was observed in photometric conditions on 1998 March
24.  The flux scale was calibrated using observations of three UKIRT
standard stars obtained on the same night.  The data were taken using
a sequence of dither motions with a typical amplitude of 15\arcsec\,
and a dwell time between dithers of 30 seconds.  The
data were linearized using an empirically measured linearity curve,
and reduced using DIMSUM\footnote{ Deep Infrared Mosaicing Software, a
package of IRAF scripts available at
ftp://iraf.noao.edu/contrib/dimsumV2}.  The total integration times
and resolutions of the resulting images are 2310~s and $1\farcs0$ at
$J$, and 2880~s and $0\farcs9$ at $K_s$.

A catalog of objects in the $K_s$-band image was obtained using
SExtractor \citep{bert96} after first geometrically
transforming the $K_s$ and $J$ images to match the $i$-band frame.
The resolution of the IR images was also degraded slightly to match
that of the $i$-band image.   Objects were detected on the $K$-band
image with the requirement that 10 contiguous pixels, covering an area
of 0.78 arcsec$^2$, must be 1.5 $\sigma$ above the background.  For
reference the 3~$\sigma$ detection limit is $K \sim 21.3$ in a 2
arcsec aperture.  All detected objects down to this limit were
inspected visually to eliminate false detections.  The catalog was
then applied to the $J$ and $i$ band images to obtain matched aperture
photometry.

\subsection{Keck Spectroscopy}

Spectroscopic observations of galaxies in a $\sim$6\arcmin\ region
around RDCS J0910+5422 were obtained using the Low Resolution Imaging
Spectrometer \citep[LRIS]{oke95} on the Keck II telescope.  Objects
were assigned slits based on their $I-K$ and $J-K$ colors.  Spectra
were obtained using the 150 l mm$^{-1}$ grating which is blazed at
7500 \AA, and covers the entire optical region, with a gradual blue
cutoff imposed by the LRIS optics at $\sim$5000~\AA.  The dispersion
of $\sim$4.8 \AA~pixel$^{-1}$ resulted in a spectral resolution of 23
\AA~as measured by the FWHM of emission lines in arc lamp spectra.
Usually each mask was observed in a series of 1800~s exposures, with
small spatial offsets along the long axis of the slitlets.  One
slitmask on the field was used to obtain spectra on 1999 March 10 with
a total exposure time of 7200~s, and a second slitmask was used on
2000 February 18 for a total of 10800~s.

The slitmask data were separated into individual slitlet spectra and
then reduced using standard longslit techniques.  A fringe frame was
constructed for each exposure from neighboring frames in the
observation sequence and then subtracted from each exposure to greatly
reduce fringing in the red.  The exposures for each slitlet were
reduced separately and then coadded.  One--dimensional spectra were
extracted for each of the targeted objects.  Wavelength calibration of
the 1-D spectra was obtained from arc lamp exposures taken immediately
after the object exposures.  A relative flux calibration was obtained
from longslit observations of the standard stars HZ 44 and G191B2B
\citep{massey88,massey90}.  While these spectra do not
straightforwardly yield an absolute flux calibration of the slit mask
data, the relative calibration of the spectral shapes is accurate.

\subsection{Chandra Imaging}

RDCS J0910+5422 was observed for a total of 200 ks in two pointings in
observation 800166 with ACIS-I.  The first pointing, exposure
ID\#2452, was performed on 2001 April 24 for 76 ks while the second
pointing, exposure ID\#2227, was conducted on 2001 April 29 for 124
ks.  For each pointing, we removed events from the level 2 event list
with a status not equal to zero and with grades one, five and seven.
In addition, we used Alexei Vikhlinin's software for removing
background events in data observed with the very faint telemetry mode.
We then cleaned bad offsets and examined the data on a chip by chip
basis, removing times when the count rate exceeded three standard
devations from the mean count rate per 3.3 second interval.  For chip
one, we specifically excluded the three brightest objects.  One of
these objects, identified with HD 237786, a G5V star, underwent a
flare in the second pointing.  For this problem, we excluded the peak
time intervals in the flare by hand.  We then cleaned each chip for
flickering pixels, i.e.\ times where a pixel had events in two
sequential 3.3 second intervals.  We finally merged the event lists
from the two pointings using the {\tt combine\_obsid} shell script
provided for this purpose.  The resulting effective exposure time for
the summed data is 163 ks.  The main reason for the relatively large
amount of time lost from the total exposure is flaring from the bright
star in the field.

\section{Results}

\subsection{Optical and Near-IR}

\begin{figure}[htb]
\plotone{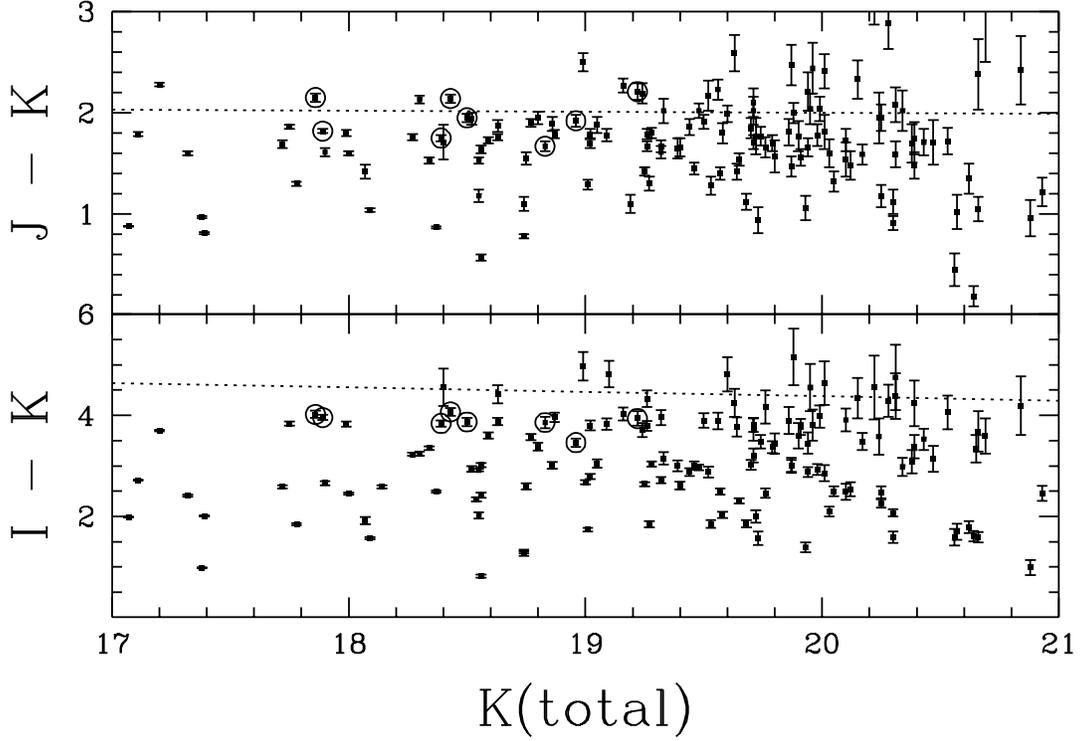}
\caption{ 
The color-magnitude diagrams from a 200\arcsec~area surrounding RDCS
J0910+5422.  Spectroscopic members are circled.  The dotted lines in
each panel show the no-evolution prediction for early-type galaxies at 
the cluster redshift, based on the observed colors of Coma
early-types \citep{sed98}.  One $\sigma$ errorbars in the color are
shown. }
\label{cmd}
\end{figure}

The presence of a group of very red galaxies at the position of RDCS
J0910+5422 is obvious in Figure~\ref{ijk}.  The spatial distribution
of the very red galaxies in Figure~\ref{ijk} is somewhat linear from
the NE to the SW; we will return to this point when discussing the
Chandra imaging.  Figure~\ref{cmd} shows the $I-K$ and $J-K$
color-magnitude diagrams for all objects in a 200 arcsec area around
RDCS J0910+5422.  A red sequence characteristic of a galaxy cluster
may be seen at $I-K \sim 4.0$ in the lower panel of Figure~\ref{cmd}.
This sequence lies some 0.5 mag to the blue of the predicted
no-evolution location for early-type galaxies at the cluster redshift.
This prediction was made using photometry of Coma galaxies as detailed
in SED98. This amount of bluing in the $I-K$ color is consistent with
the color change due to passive evolution of a single age $Z_\odot$
stellar population formed in a 0.1 Gyr burst at $z_f = 3$ (using the
GISSEL models of \citet{bc00}).  There is an indication that the red
sequence in RDCS J0910+5422 has a flatter slope relative to the Coma
sequence, but this result is very uncertain.  The slope difference is
not likely to be due to differences in the photometry of the galaxies
in Coma and RDCS J0910+5422---e.g., apertures of the same physical
size were used on both clusters.  A similarly flatter slope was
tentatively found in RDCS J0848+4453 at $z=1.27$ \citep{dokkum01b}.
The observed scatter in the $I-J$ colors of the member galaxies is
0.09 mag and the measurement error is 0.08 mag, indicating a very
small amount of intrinsic scatter, $\sim$0.04, in the rest frame $\sim
U-V$ colors.
  
The optical spectra for the 9 member galaxies are presented in
Figure~\ref{optspec}.  Redshifts were calculated both by visual
identification of emission and absorption features and by
cross-correlating the spectra with an E template from \citet{kinney96}
using the IRAF Package RVSAO/XCSAO \citep{kurtz91} and are listed in
Table~\ref{table1}.  The redshift measurements primarily are based on
major features such as Ca II H+K and OII$\lambda3727$, and is also
sensitive to spectral breaks such as D4000 and B2900. Spectra were
obtained for a total of 15 color-selected targets; of these redshifts
were determined for 13 and 8 are cluster members (one serendipitous
spectrum is a member).  The locations of the member galaxies are shown
on the color composite image in Figure~\ref{ijk}.  ID~\#161, which
lies outside our $K_s$ image, was discovered serendipitously in an
LRIS mask to be at the cluster redshift and has relatively strong
[OII].  Two objects (ID~\#23 and 57) show weak [OII] emission, while
the remaining member galaxies have spectra typical of early-type
galaxies in the present epoch, albeit with smaller D4000.

\subsection{X-ray}

To measure the X-ray flux in the ACIS-I image, we centered a circular
aperture with a radius of 100\arcsec\ at 09h10m44\farcs9,
+54d22m08\farcs9 (J2000).  The position was chosen by the flux
weighted centroid of all events at 0.5--2.0 keV within 20\arcsec\ of
the visual center of the cluster.  From this circular aperture, we
excluded seven point sources.  Each point source was identified from
the smoothed contour map overlaid on the composite optical and near
infrared image.  We then fit an elliptical $\beta$ model and a
constant background to the events within the 0.5--2.0 keV map.  We
used a map binned into 1\arcsec\ pixels and fit the model using the
CIAO package Sherpa \citep{free00} with the Cash statistic \citep{cash79}.
We first explored the parameter space with 3000 Monte Carlo samples,
and then refined the best fitting Monte Carlo result.  Our best
fitting model has a core radius of 19\farcs 4 $\pm$ 0.6 and a $\beta =
0.887^{+0.028}_{-0.026}$.  Our best-fitting model was mildy
elliptical, with an ellipticity of $0.045^{+0.048}_{-0.045}$.  At the
redshift of the cluster, the core radius would be 171.1 $\pm$ 5.3 kpc.
These values are typical of low redshift clusters.

For an estimate of the background spectrum, we chose three separate
regions.  Each background region was visually inspected which resulted
in the removal of bright point sources.  We fit all three regions
jointly, using the program {\tt XSPEC} \citep{arnaud96}, with a two
component model (refered to as a background model in the program {\tt
XSPEC}) consisting of a powerlaw, not convolved with the telescope
effective area, and a Gaussian for the 2.1 keV Au emission line.  Each
region had a seperate normalization for the two components.  We used
separate response matrices for each region, and we generated these
using Alexei Vikhlinin's calcarf/calcrmf tools.

\begin{figure}[htb]
\plotone{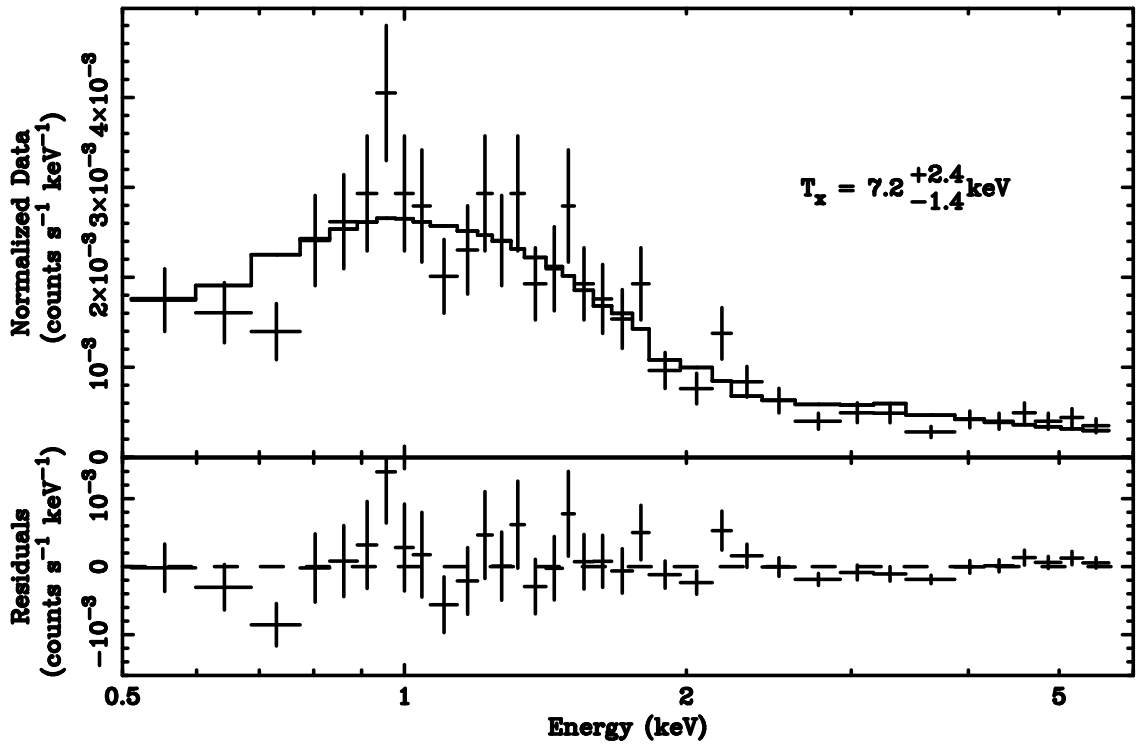}
\caption{ 
The folded ACIS data within a $\sim$38 arcsec aperture centered on the 
X-ray emission plotted with the unfolded model.  The model has been
grouped into bins containing at least 20 events.}
\label{acis_spec}
\end{figure}

We extracted a spectrum of the cluster using an elliptical aperture.
The semi-major axis of the ellipse was twice the core-radius of the
cluster.  We fixed the background to the values from the best fitting
model above, with the normalizations rescaled by the relative area in
the aperture.  Freezing the background and the absorption by galactic
hydrogen at $2\times 10^{20}$ cm$^{-2}$, obtained from the
100~\micron\ maps of \citet{schlegel98}, we fit a Raymond-Smith
spectrum using the Cash statistic to the spectrum shown in
Figure~\ref{acis_spec} in the source aperture. We found a best fitting
temperature of kT$ = 7.2^{+2.4}_{-1.2}$ keV.  The best fitting flux in
the {\tt ROSAT} band of 0.5-2.0 keV was $1.06^{+0.07}_{-0.06}\times
10^{-14}\ {\rm ergs\ cm^{-2}\ s^{-1}}$ within the 38\farcs 8 aperture.
The flux as measured in the RDCS in the same band from Rosat data was
$2.0 \times 10^{-14}\ {\rm ergs\ cm^{-2}\ s^{-1}}$.  Most of the
difference between the Rosat and Chandra fluxes is due to the
exclusion of the point sources detected in the ACIS image from the
latter flux.  The total bolometric luminosity of RDCS J0910+5422 is
$2.48^{+0.33}_{-0.26}\times 10^{44}{\rm ergs\ s^{-1}}$ when integrated
over the whole of the $\beta$ model.  The errors on the temperature,
flux, and luminosity were determined using 2000 iterations of the {\tt
fakeit} command in {\tt XSPEC}.  Unfortunately we are unable to
determine the metallicity of the ICM to see if the canonical 1/3
$Z_\odot$ seen in clusters up to $z \sim 0.8$ \citep{don99,ml97}
continues beyond $z = 1$.

A total mass can be estimated from the $T_x$, assuming an isothermal
sphere and extrapolating the X-ray emission to $r = 1$ Mpc using the
best fit profile \citep{borg99,horner99}.  The total mass of
RX~J0910+5422 derived from the new X-ray data is $7.0^{+2.3}_{-1.1}
\times 10^{14} M_{\sun} h_{65}^{-1}$ within $r = 1$ Mpc.  Recent
results obtained with Chandra and XMM on the ICM in lower-redshift
clusters indicate that, apart from drops in the central regions due to
cooling flows, the temperature profiles are fairly constant out to
large radii, implying that mass estimates based on the assumption of
isothermality are reasonable \citep{borg01b}.

Several point sources were detected in the vicinity of the extended
X-ray emission of the cluster, demonstrating the importance of
high spatial resolution when attempting to accurately measure the
properties of the ICM in high-$z$ clusters.  One of the fainter point
sources is associated with a cluster member, ID\#23, which has weak
emission lines and a relatively blue continuum.  This object has $L_x
= 5.1 \times 10^{42}{\rm ergs\ s^{-1}}$ in the [0.5--10] keV band.  As
seen in Figure~\ref{ijk} there is a close neighbor with which ID\#23
could be interacting; higher resolution imaging will be useful to
answer this question.  ID\#23 is probably a low luminosity AGN, though
we did not find evidence of the NeV line at $\lambda$3426 in our LRIS
spectrum which is usually detected in such objects.  A second much
brighter X-ray point source ($F_x = 0.73 \times 10^{-15}\ {\rm ergs\
cm^{-2}\ s^{-1}} $) just to the northeast of cluster center is
associated with the $K$-band object ID\#21 which has the colors ($J-K
= 1.86$ and $I-K = 3.85$) of a cluster galaxy but for which we have no
spectrum. A third point source, ID\#61, is also of interest because
it is a hard X-ray source and a photometric member. Such galaxies are of
particular value as they may represent a source of early heating
during the formation of the ICM.

Though the X-ray morphology largely appears to be that of a relaxed
system, there are indications that this cluster is still forming.
Indeed given its high redshift and the expectations of cluster
building which should occur at these redshifts according to
$\Lambda$CDM simulations, evidence of e.g.\ mergers are to be
expected.  As shown in Figure~\ref{shock}, there is some evidence in
the ACIS data for temperature structure or merging in the ICM.  The soft
component dominates the central area of the X-ray emission, while to
the south there is a harder component.  Such a temperature
distribution in the ICM could be due to an infalling group, or to mass
streaming in along a filament.  Indeed, the spatial distribution of
the galaxies in the cluster gives the impression of filamentary
structure, reminiscent of that seen in CDM simulations of cluster
formation \citep{frenk99}.

\section{Summary}

\subsection{The cluster galaxy population}

The spectrophotometric properties of all the known members in RDCS
J0910+5422 are summarized in Table 1.  Two of the member galaxies show
signs of current star formation, and one of these may be an AGN due to
its X-ray emission.  The spectra and colors of 6 of the spectroscopic
members are broadly similar to those of a passively evolving
elliptical galaxy formed at $z \sim 3$.  But from these data alone, it
is unclear if the luminous galaxies in the red sequence formed via
hierarchical merging at $z < 3$ or as a single object at $z > 3$.  Age
dating the spectra could provide additional information on the
formation epoch of the galaxies; but such detailed modelling is
burdened with its own uncertainities due to the age--metallicity
degeneracy.  Progress on the issue of the assembly of early-type
galaxies in clusters is most likely to occur with determining the
morphologies of the member galaxies using high resolution HST imaging
on a large sample within RDCS J0910+5422, and in other similarly
high-$z$ clusters.  

\subsection{ICM}

\begin{figure}[htb]
\plotone{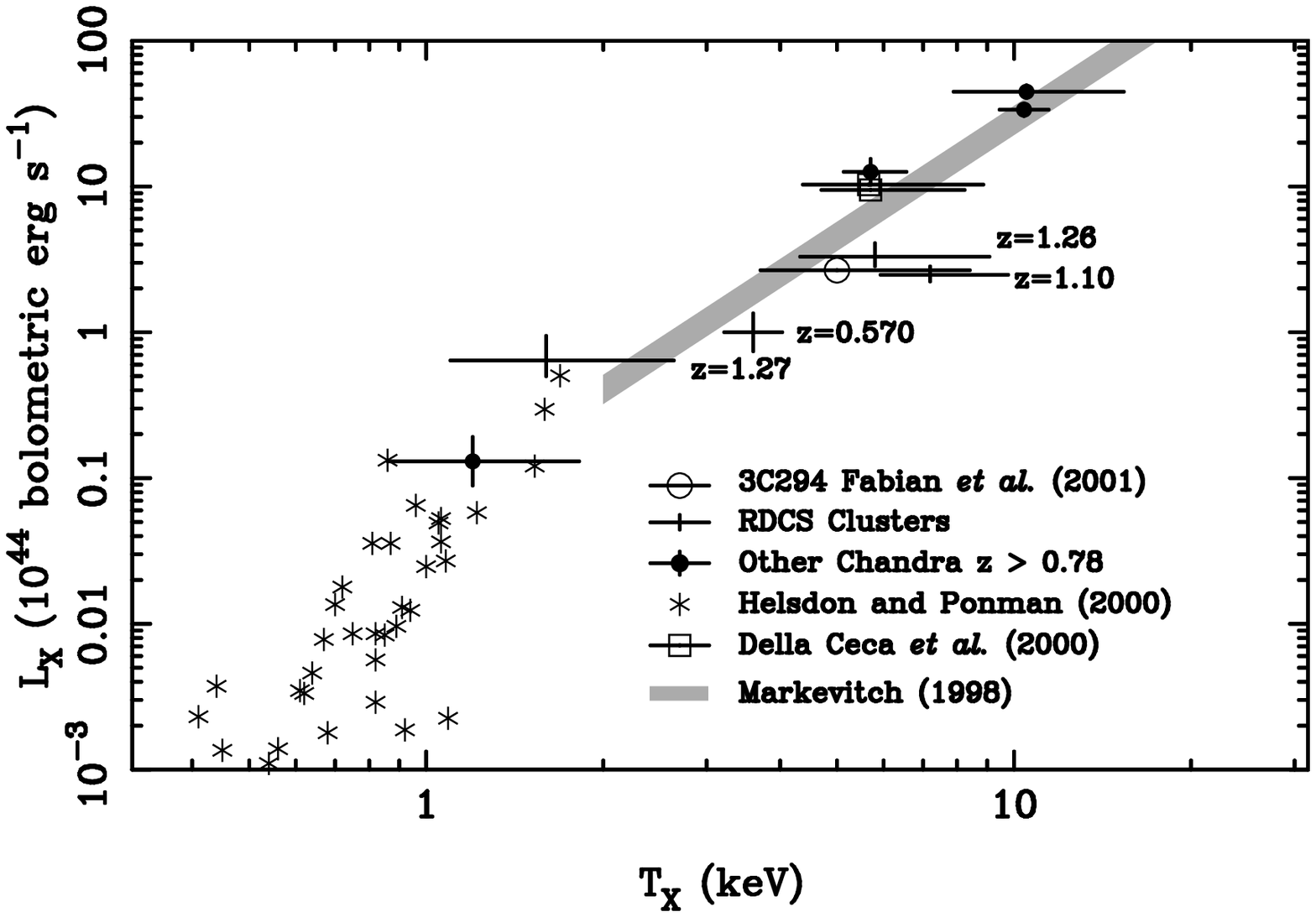}
\caption{The X-ray luminosity vs X-ray temperature for various cluster
samples as detailed in the legend, with $H_0=65$ km s$^{-1}$
Mpc$^{-1}$, $\Omega_m = 0.3$, and $\Omega_\Lambda = 0.7$. }
\label{lxtx}
\end{figure}

The Chandra data conclusively show the presence of hot gas trapped in
the potential well of a massive cluster.  This is the third instance
in the RDCS of a cluster at $z > 1$ with a well-defined ICM.  The ACIS
spectra yield a $kT = 7.2^{+2.2}_{-1.4}$ keV, and a total $L_x =
2.48^{+0.33}_{-0.26}\times 10^{44}{\rm ergs\ s^{-1}}$, both near the
values for an $L^\ast$ cluster in the present epoch.  Furthermore, the
position of RDCS J0910+5422 in a plot of $L_x$ vs $T_x$, shown in
Figure~\ref{lxtx}, shows little if any evolution in the $L-T$ relation
at $z > 1$, in keeping with the results of Borgani et al.\ (2001).
Along with the well-defined red envelope, these properties indicate
that RDCS J0910+5422 is another example of a massive cluster with an
old galaxy population and a hot ICM already in place at $z > 1$.

\acknowledgments

The authors thank the staffs of Keck and Palomar Observatories, and
the builders of the Chandra X-ray Observatory for providing the means
with which we obtained our data.  We also thank Paolo Tozzi for
assistance with the reductions of the Chandra data, and Tom Jarrett
for assistance with the use of PFIRCAM at Palomar.  Support for SAS
came from NASA/LTSA grant NAG5-8430 and for BPH from Chandra grant
GO1-2141A.  Part of the observational material presented here was
obtained at the W.\ M.\ Keck Observatory, which is a scientific
partnership between the University of California and the California
Institute of Technology, made possible by a generous gift of the W.\
M.\ Keck Foundation.  The work by SAS and BPH at LLNL was performed
under the auspices of the U.S. Department of Energy under Contract
No.\ W-7405-ENG-48.  Portions of this work were carried out by the Jet
Propulsion Laboratory, California Institute of Technology, under a
contract with NASA.

\begin{figure}[htb]
%\plotone{figure1.ps}
\caption{ A color composite made from the $iJK$ band images with N up
and E to the left.  The field size is 150\arcsec. 
Spectroscopically-confirmed members at $1.11$ are marked by boxes;
probable cluster members selected by their $I-K$ and $J-K$ colors are
circled. The ID numbers of the members are \# 23, 57, 37, 54, 24, 35,
38, and 68 from North to South; \#161 lies outside the $K$-band field. }
\label{ijk}
\end{figure}

\begin{figure}[htb]
\plotone{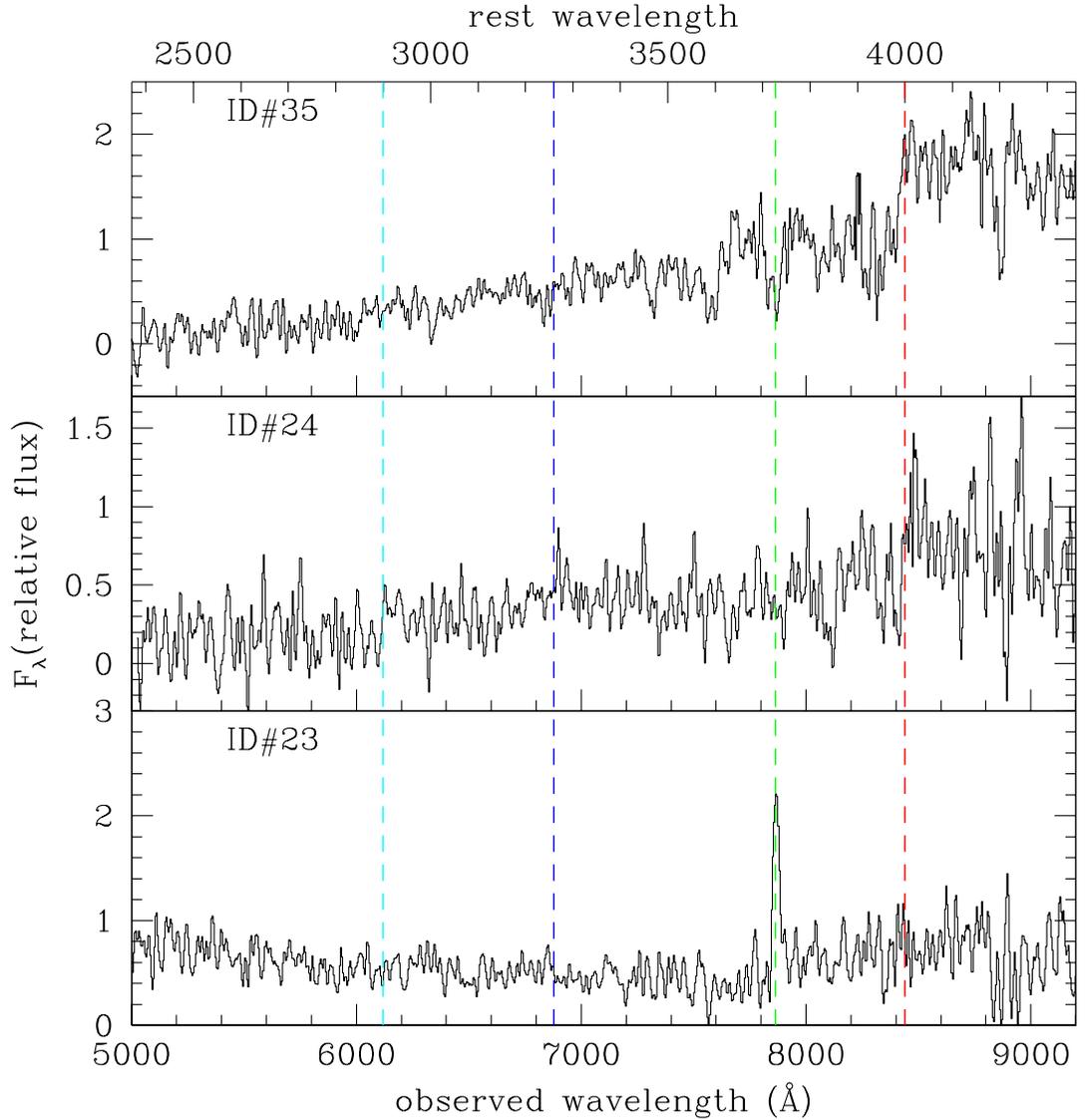}
\caption{LRIS spectra of cluster members obtained at Keck.  The rest
frame wavelengths are shown along the top for $z = 1.10$ and the
associated positions of major spectral features (from left to right:
B2900, B3260, $\lambda$3727, and D4000) are marked by vertical
dashed lines.  The flux calibration is relative. }
\label{optspec}
\end{figure}

\begin{figure}[htb]
\figurenum{3}
\plotone{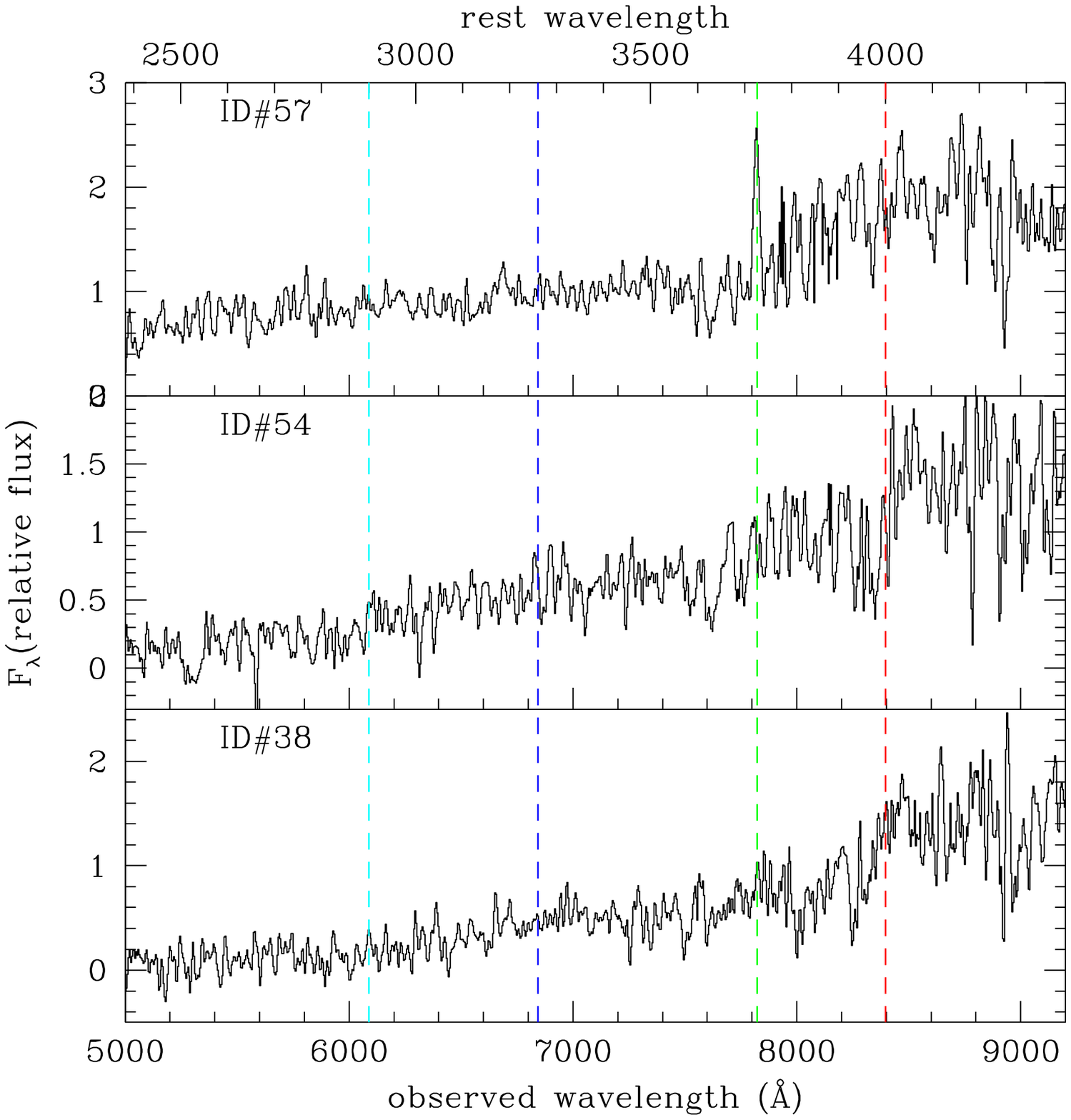}
\caption{continued}
\end{figure}

\begin{figure}[htb]
\figurenum{3}
\plotone{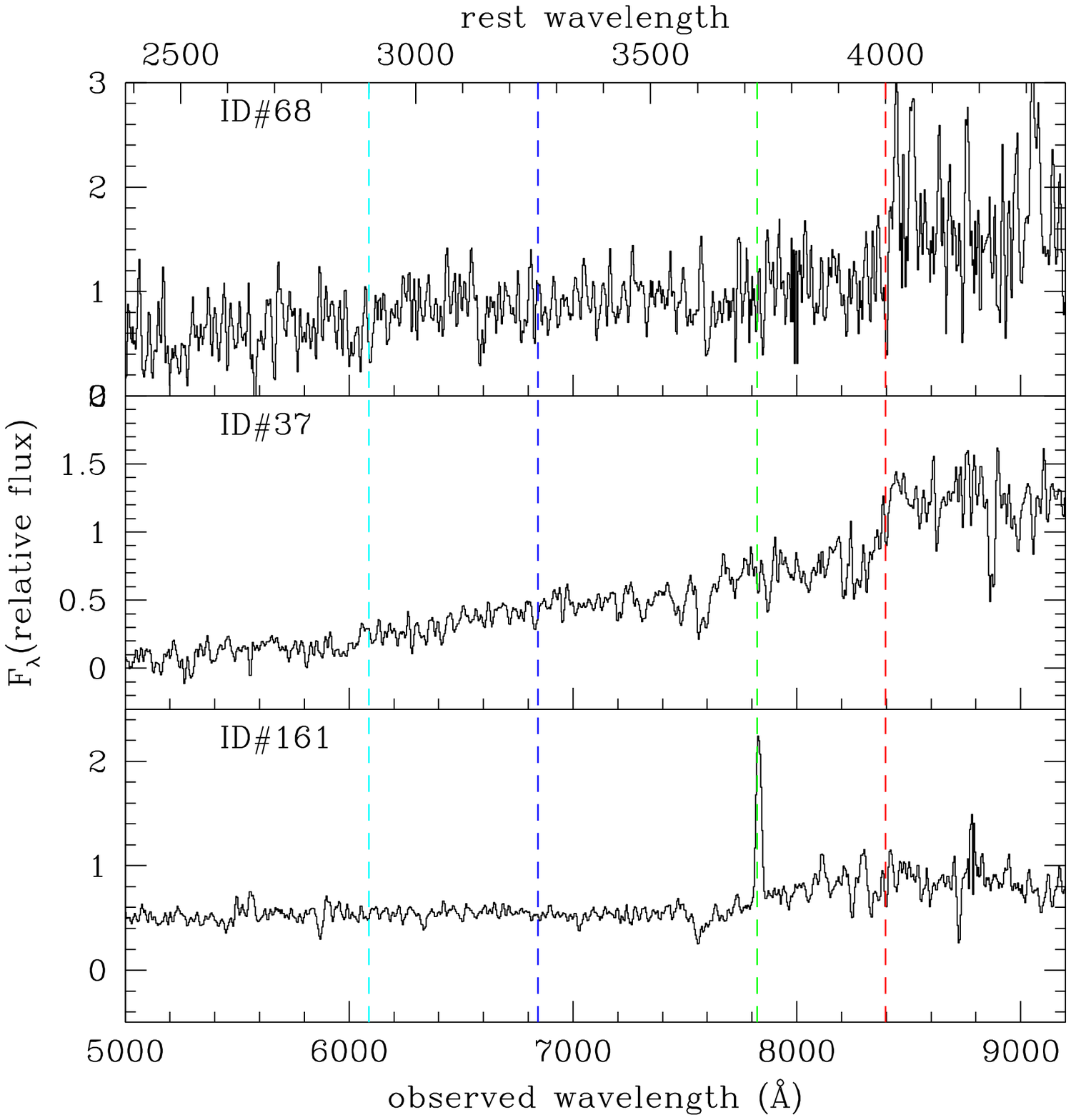}
\caption{continued}
\end{figure}

\begin{figure}[htb]
%\plotone{figure4.ps}
\caption{$K_s$ band image of RDCS J0910+5422. The X-ray contours of
the ACIS data are from the 0.5--2.0 keV events smoothed with a
5\arcsec\ FWHM Gaussian.  The X-ray emission is not centered on an
obvious brightest cluster galaxy and has a shape slightly elongated in
the same direction as that of the member galaxy distribution.  The
bright X-ray point source, ID\#21, just northeast of the cluster is associated
with a photometric cluster member.  The fainter X-ray point source
even further to the northeast (located at R.A.~$=$9:10:48.34,
Dec$=$54:22:29) is associated with galaxy ID\#23, a spectroscopic
cluster member. No spectral information is available for ID\#27, which
does not have the colors expected of a cluster member. }
\label{kx}
\end{figure}

\begin{figure}[htb]
%\plotone{figure6.ps}
\caption{The left panel shows the 0.5--2.0 keV emission and the right
panel shows the 2.0--6.0 keV emission.  The smoothing scale is FWHM $=
4.7\arcsec$.}
\label{shock}
\end{figure}

\begin{deluxetable}{ccccccc}
\tablenum{1}
\tablecaption{Spectrophotometric properties of galaxies in RDCS J0910+5422.}
\tablehead{
\colhead{ID} & \colhead{R.A.} & \colhead{Dec.} & 
\colhead{$K$} & \colhead{$J-K$} & \colhead{$I-K$} & 
\colhead{$z$}  
}
\startdata

23 & 9:10:48.34 & 54:22:29 & 17.86 & 2.15 & 4.02 & 1.1108 \\
24 & 9:10:44.99 & 54:22:02 & 17.89 & 1.82 & 3.96 & 1.1075 \\
35 & 9:10:44.88 & 54:21:59 & 18.39 & 1.75 & 3.84 & 1.1196 \\
37 & 9:10:46.27 & 54:22:11 & 18.43 & 2.14 & 4.07 & 1.105~  \\
38 & 9:10:42.85 & 54:21:43 & 18.50 & 1.95 & 3.87 & 1.0951 \\
54 & 9:10:45.34 & 54:22:04 & 18.83 & 1.67 & 3.86 & 1.0997 \\
57 & 9:10:48.30 & 54:22:24 & 18.96 & 1.92 & 3.46 & 1.0989 \\
68 & 9:10:50.15 & 54:21:03 & 19.22 & 2.21 & 3.94 & 1.107~  \\
161\tablenotemark{a} & 9:10:30.08 & 54:18:45 & \nodata & \nodata & \nodata & 1.1136 \\
\enddata
\tablenotetext{a}{Object outside area covered by our $K$ band image.}
\label{table1}
\end{deluxetable}

\end{document}